\documentclass{aa}  

\usepackage{graphicx}
\usepackage[dvipsnames]{xcolor}
\usepackage[normalem]{ulem}
\usepackage{txfonts}
\usepackage{mathtools}
\usepackage{booktabs}
\usepackage{hyperref}
\hypersetup{
    colorlinks=true,
    linkcolor=blue,
    urlcolor=blue,
    pdftitle={LyαNNA II: Field-level inference with noisy Lyα forest spectra},
    pdfauthor={Parth Nayak and Michael Walther and Daniel Gruen}
}

\newcommand{\python}{\textsc{Python}}
\newcommand{\optuna}{\textsc{Optuna}}
\newcommand{\keras}{\textsc{Keras}}
\newcommand{\tensorflow}{\textsc{TensorFlow}}

\newcommand{\nyx}{\textsc{Nyx}}
\newcommand{\synth}{\textsc{SynTH}}
\newcommand{\lyanna}{\textsc{Ly$\alpha$NNA}}
\newcommand{\lya}{Ly$\alpha$}

\newcommand{\nsansa}{\textsc{nSansa}}
\newcommand{\plain}[1]{\mathrm{#1}}
\newcommand{\define}{\coloneqq}
\newcommand{\TPS}{\mathbb{P}}
\newcommand{\Det}[1]{\vert #1 \vert}
\newcommand{\LogDeterminantSigma}{\log \vert \tilde{\mathbf{\Sigma}} \vert}
\newcommand{\score}{\mathcal{L}^*_\plain{val}}
\newcommand{\delchisq}{\delta \chi^2_\plain{r}}

\begin{document}

   \title{{\lyanna} II: Field-level inference with noisy {\lya} forest spectra}

   \author{
        Parth Nayak\inst{1,2}\thanks{\email{parth3e8@gmail.com}} 
        \and Michael Walther\inst{1,3} 
        \and Daniel Gruen\inst{1,2,3}
        }

   \institute{
        University Observatory, Faculty of Physics, Ludwig-Maximilians-Universit\"at, Scheinerstr. 1, 81679 Munich, Germany
        \and Munich Center for Machine Learning (MCML), Oettingenstr. 67, 80538 Munich, Germany
        \and Excellence Cluster ORIGINS, Boltzmannstr. 2, 85748 Garching, Germany
        }

   \date{Received XXXX; accepted YYYY}

 \abstract
    {
        Deep learning (DL) has been shown to outperform traditional, human-defined summary statistics of the {\lya} forest in constraining key astrophysical and cosmological parameters owing to its ability to tap into the realm of non-Gaussian information. An understanding of the impact of nuisance effects such as noise on such field-level frameworks, however, still remains elusive. 
        
        In this work we conduct a systematic investigation into the efficacy of DL inference from noisy {\lya} forest spectra. Building upon our previous, proof-of-concept framework \citep{Nayak_2024_LyaNNA} for pure spectra, we constructed and trained a ResNet neural network using labeled mock data from hydrodynamical simulations with a range of noise levels to optimally compress noisy spectra into a novel summary statistic that is exclusively sensitive to the power-law temperature-density relation of the intergalactic medium. 
        We fit a Gaussian mixture surrogate with 23 components through our labels and summaries to estimate the joint data-parameter distribution for likelihood free inference, in addition to performing inference with a Gaussian likelihood. The posterior contours in the two cases agree well with each other.
        
        We compared the precision and accuracy of our posterior constraints with a combination of two human defined summaries (the 1D power spectrum and PDF of the {\lya} transmission) that have been corrected for noise, over a wide range of continuum-to-noise ratios (CNR) in the likelihood case. We found a gain in precision in terms of posterior contour area with our pipeline over the said combination of 65\% (at a CNR of 20 per 6 km/s) to 112\% (at 200 per 6 km/s). While the improvement in posterior precision is not as large as in the noiseless case, these results indicate that DL still remains a powerful tool for inference even with noisy, real-world datasets. 
    }
   

   \maketitle

\section{Introduction}

  The interplay of the continua of quasars' emission %
  and the intergalactic medium (IGM) imprints a unique and powerful absorption feature on their observed spectra called the {\lya} forest \citep{Lynds_1971_LyaF}. %
  The term refers to a dense cluster of resonant {\lya} absorption lines of Hydrogen blueward of a quasar's %
  observed {\lya} emission peak, spread over a wide range of wavelengths as a result of the expansion of the universe which causes the cosmological redshift. Since Hydrogen is the most abundant element in the intergalactic gas,
  the {\lya} forest provides a continuous one dimensional probe of the IGM along the quasar line of sight. %

  Individual high resolution spectroscopic observations of quasars using state-of-the-art instruments such as VLT/UVES \citep[e.g.,][]{SQUAD_2019MNRAS.482.3458M}, VLT/XSHOOTER \citep[e.g.][]{XQ100_2016A&A...594A..91L}
  and Keck/HIRES \citep[e.g.,][]{KODIAQ2_2017AJ....154..114O}, %
  as well as large cosmological spectroscopic surveys such as the extended Baryon Oscillation Spectroscopic Survey (eBOSS, \citealt{eBOSS_Main_2013AJ....145...10D}) and most recently the Dark Energy Spectroscopic Instrument (DESI, \citealt{DESI_OVERVIEW}) %
  have opened a spectroscopic treasure trove of cosmological information such as submegaparsec scale nonlinearities 
  in the cosmic matter distribution, and the thermal and ionization state of the IGM. These {\lya} forest observations have been widely used to constrain the thermal history 
  of the IGM and the epoch of reionization (e.g., \citealt{Becker_Inference_curvature_2011MNRAS.410.1096B, Walther_IGM_2019ApJ...872...13W, Boera_Inference_P1D_2019ApJ...872..101B, Gaikwad_IGM_2021MNRAS.506.4389G, Bosman_Reion_2022MNRAS.514...55B}), the baryon acoustic oscillation (BAO) scale (e.g., \citealt{Slosar_BAO_2013JCAP...04..026S, Busca_BAO_2013A&A...552A..96B, eBOSS_BAO_Lya_2020ApJ...901..153D, DESI_DR2_BAO_2025arXiv250314739D}), and the properties of dark matter (e.g., \citealt{Viel_DM_2005PhRvD..71f3534V, Irsic_FDM_2017PhRvL.119c1302I, Armengaud_DM_2017MNRAS.471.4606A, Kobayashi_DM_2017PhRvD..96l3514K, Rogers_DM_2021PhRvL.126g1302R}). When combined with external probes such as the cosmic microwave background (CMB, e.g., \citealt{Planck2018_2020A&A...641A...6P}), they provide valuable insights into cosmic inflation and the mass of neutrinos (e.g., \citealt{Seljak_Cosmo_2006JCAP...10..014S, Palanque-Delabrouille_Neutrino_2015JCAP...11..011P, Yeche_Neutrino_2017JCAP...06..047Y, Palanque-Delabrouille_Neutrino_2020JCAP...04..038P, Sarkar_Neutrino_2024JCAP...09..003S, Ivanov_Neutrino_2025PhRvL.134i1001I}).

  The classical way of gaining insights from the {\lya} forest is through human defined summary statistics. For instance, the 1D power spectrum, a measure of the absorption structure (and hence the underlying IGM) across length scales, is one of the most widely used summary statistics in literature for parameter inference (e.g., \citealt{Croft_P1D_1998ApJ...495...44C, Theuns_P1D_2000MNRAS.315..600T, McDonald_FPDF_2000ApJ...543....1M, Chabanier_P1D_2019JCAP...07..017C, Walther_IGM_2019ApJ...872...13W, Boera_Inference_P1D_2019ApJ...872..101B, DESI_P1D_FFT_2025arXiv250509493R, DESI_P1D_QMLE_2025arXiv250507974K}). Other examples of summary statistics include the probability density function (PDF) of the {\lya} transmission (e.g., \citealt{Theuns_P1D_2000MNRAS.315..600T, McDonald_FPDF_2000ApJ...543....1M, Bolton_FPDF_2008MNRAS.386.1131B, Viel_Inference_FPDF_2009MNRAS.399L..39V, Lee_FPDF_2015ApJ...799..196L}), curvature statistics (e.g., \citealt{Becker_Inference_curvature_2011MNRAS.410.1096B, Boera_Inference_curvature_2014MNRAS.441.1916B}), and wavelet statistics (e.g., \citealt{Meiskin_wavelets_2000MNRAS.314..566M, Theuns_wavelets_2000MNRAS.317..989T, Zaldarriaga_wavelets_2002ApJ...564..153Z, Lidz_Wavelets_2010ApJ...718..199L, Wolfson_Inference_Wavelet_2021MNRAS.508.5493W, Tohfa_Wavelet_Scattering_2024PhRvL.132w1002T}). %
  However, those summary statistics do not efficiently compress the information of the non-Gaussian {\lya} forest transmission
  field, restricting the constraining power of the full data. %

  In %
  recent years, machine learning methods are gaining popularity for the task of parameter inference in cosmology due to their ability to optimally extract and compress relevant non-Gaussian features from the underlying field. Presently, weak gravitational lensing is one of the most common cosmological probes with machine learning inference applications in the
  literature (e.g., \citealt{Gupta_PhysRevD.97.103515, Fluri_KiDS450_PhysRevD.100.063514, Ribli_2019MNRAS.490.1843R, Lu_2022MNRAS.511.1518L, Fluri_KiDS1000_2022PhRvD.105h3518F, LucasMakinen_2025JCAP...01..095L, DESY3_NN_2025MNRAS.536.1303J}).
  For the {\lya} forest, the avenue of field-level neural network inference remains relatively unexplored with only a handful of works (e.g., \citealt{Nayak_2024_LyaNNA, Maitra_Lya_IMNN_2024A&A...690A.154M, Nasir_2024MNRAS.534.1299N}). 
  In \citet[][hereafter N24]{Nayak_2024_LyaNNA}, we demonstrated the potential of deep learning for parameter inference from the {\lya} forest spectra for the first time. The field-level neural network (NN) framework we presented therein (called {\lyanna}) surpasses the traditional, human defined summaries in the constraining power significantly, as substantiated by \citet[][hereafter C25]{Chang_2025_Summaries}.

  The factors-of-a-few improvement by using {\lyanna} as presented in N24 and C25, %
  however, is a proof-of-concept, in the regime with no realistic nuisances modeled on top of the pure simulations of the {\lya} forest. In real observations, a plethora of astrophysical and instrumental systematics go hand in hand with the cosmological signal of interest. These include, but are not limited to, instrumental noise and spectral resolution, sky lines, uncertainties in the estimation of the intrinsic quasar continua, high column density systems (HCDs), and intergalactic metal absorption \citep[see, e.g.,][for a discussion of systematics in the context of DESI]{DESI_P1D_Val_2025arXiv250913593K}. Each of these effects %
  acts as a contaminant to the {\lya} forest, potentially %
  compromising the full constraining power of the signal. They may add uncertainty to the data (thereby reducing inference precision) and they may also bias the summary vectors (and thus the posterior distributions). With a theoretical understanding of the impact of some of these effects (e.g. noise, resolution) on some of the traditional summaries, their interference may be corrected, allowing us to make the most of our contaminated datasets. For instance, noise in the {\lya} forest spectra adds a measurable bias to the power spectrum, and spectral resolution dampens the power on small scales with a known kernel (see, e.g., \citealt{Walther_P1D_2018ApJ...852...22W}). However, for inference with machine-learned summaries such as {\lyanna}, we lack such understanding. %

  In this work, we explore field-level inference with deep learning using contaminated {\lya} forest datasets. In particular, we analyze the {\lyanna} framework in the presence of noise and limited spectral resolution---two of the strongest contaminants in  spectroscopic  datasets 
  ---and investigate whether the same amount of improvement may be gained from such observations as for pure simulations while inferring the amplitude and slope of the power-law temperature density relation of the IGM. We retrained the pipeline described in N24 with a range of signal-to-noise ratio values in the training data and an adapted architecture to better process them. After an extensive hyperparameter search rooted in Bayesian optimization, we trained a committee 
  of 2200 %
  NNs with the best performing architecture for ensemble learning \citep{Dietterich_Ensemble_Learning}. We built an emulator of our NN compressed summary vector for likelihood-based inference as well as a Gaussian mixture model (GMM) surrogate of the joint density of the NN summary and the true parameter values for likelihood-free inference (LFI, see e.g. \citealt{Grazian_2019arXiv190902736G} for a review)
  and made comparisons between them. For a Gaussian likelihood case, we compared the inference outcome of our machinery to that of the 1D power spectrum and the transmission PDF to assess the improvement of NN for noisy spectra. 

  We organize this paper as follows. In Section~\ref{sec:sim} we describe the hydrodynamical simulations and the mock datasets we used for training our machinery, including the realistic effects modeled. In Section~\ref{sec:machinery} we present a detailed overview of the NN machinery used in this work. In Section~\ref{sec:inference-method} we describe our approach to Bayesian inference with our machinery including LFI. We show and discuss our results of using this framework in Section~\ref{sec:results} with a comparison to some of the traditional statistics. We conclude with a summary of this work and an outlook in Section~\ref{sec:conclusion}.
  
\section{Simulations}\label{sec:sim}

    \subsection{Hydrodynamic simulations and mock spectra}\label{sub:hydro-sim}
        For creating the mock {\lya} forest spectra used in this work, we used a {\nyx} cosmological hydrodynamic simulation snapshot from the Lyssa suite described in %
        \citet{Walther_Lyssa_2025JCAP...05..099W} generated primarily for eBOSS and %
        DESI {\lya} forest analyses. {\nyx} 
        is a hydrodynamics code that simulates 
        baryons as an ideal fluid on an Eulerian grid, gravitationally coupled with N-body dark matter particles. In our simulation, initial conditions are generated at $z=99$ using second order Lagrangian perturbation theory. The gas is evolved by solving the %
        Euler equations using a second-order accurate scheme. {\nyx} models all the key physical processes affecting the IGM and hence the {\lya} forest such as collisional and photoionization, recombination, photoheating and cooling of the gas that has a primordial composition of H and He (for more details see \citealt{Almgren_Nyx_2013ApJ...765...39A} and \citealt{Lukic_Nyx_Lya_2015MNRAS.446.3697L}). In our simulation, a spatially flat, time-varying ionizing ultraviolet background (UVB) was applied following the late reionization model of \citet{Onorbe_UVB_2017ApJ...837..106O}, and the radiative and dielectric recombination, collisional ionization and cooling rates from \citet{Lukic_Nyx_Lya_2015MNRAS.446.3697L} were used. Our snapshot has a comoving side length of 80 Mpc$/h$ and contains $4096^3$ volumetric cells (aka voxels) and dark matter particles. This snapshot is the ``fiducial'' run from \citet{Walther_Lyssa_2025JCAP...05..099W} at $z=2.2$  %
        with the cosmological parameters based on %
        \citet{Planck2018_2020A&A...641A...6P}, namely, $h=0.6732$, $\Omega_\mathrm{m}=0.3144$, $\Omega_\mathrm{b}=0.0494$, $A_\mathrm{s} = 2.101\times10^{-9}$, $n_\mathrm{s} = 0.9660$. 

        The diffuse IGM  that constitutes the bulk of baryonic gas in the universe follows a tight power law
        temperature-density relation (TDR) characterized by
        \begin{equation}\label{eqn:tdr}
            T = T_0 (\rho_\mathrm{b}/\bar{\rho}_\mathrm{b})^{\gamma - 1},
        \end{equation}
        where $T_0$ is a temperature at the cosmic mean gas density $\bar{\rho}_\mathrm{b}$ and $\gamma$ is the adiabatic power-law index \citep{Hui_Gnedin_1997MNRAS.292...27H}. The fiducial values of these two parameters in our snapshot are $T_0 = 10765$ K and $\gamma = 1.57$ (estimated by a linear least-squares fitting in the range $-0.5<\log_{10}(\rho_\mathrm{b}/\bar{\rho}_\mathrm{b})<0.5$ and $\log_{10}(T/\mathrm{K})<4$).

        \begin{figure}
            \centering
            \includegraphics[width=\linewidth]{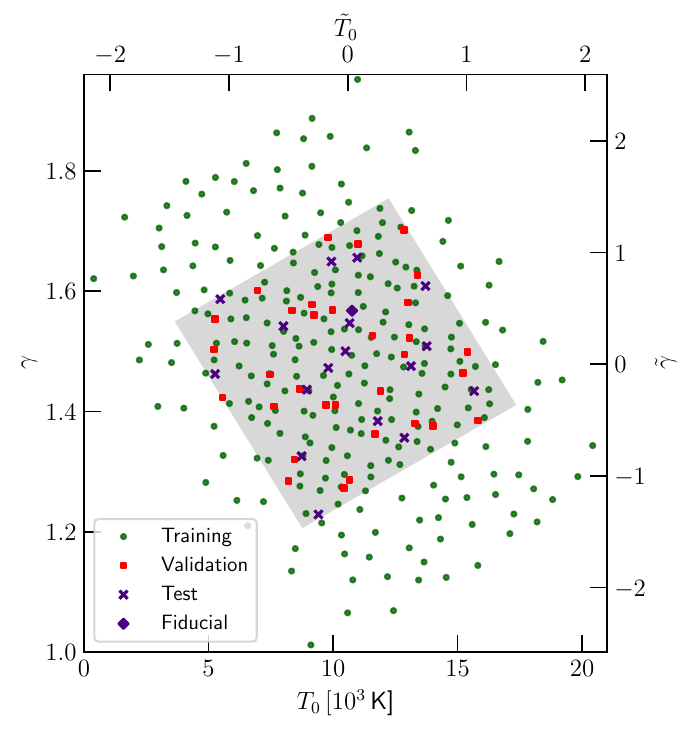}
            \caption{\small The sample of training, validation, and test labels in our mock dataset along with the fiducial TDR model in the $(T_0,\gamma)$ as well as the rescaled $(\tilde{T}_0, \tilde{\gamma})$ space. The gray shaded region indicates our prior for the likelihood analysis as well as for the density estimation likelihood free inference. The exact sampling strategy is described in Appendix~\ref{app:sampling}.
            } 
            \label{fig:param-space-sample-Planck18}
        \end{figure}
        We used an original {\python} package called {\synth}\footnote{Synthesis of Transmission from Hydrodynamical simulations; \href{https://github.com/par-nay/synth}{https://github.com/par-nay/synth}.} for the generation of mock spectra from the simulation snapshot that is different from the pipelines used by \citet{Lukic_Nyx_Lya_2015MNRAS.446.3697L} and \citet{Walther_Lyssa_2025JCAP...05..099W}.
        We rescaled the gas temperatures at fixed densities as a post-processing step to populate the $(T_0, \gamma)$ parameter space---and account for the scatter off the power law relation---for creating labeled datasets for our supervised learning framework, same as N24. We also applied a rescaling to our parameter labels $T_0\to\tilde{T}_0$ and $\gamma\to\tilde{\gamma}$, similar to N24---such that $\tilde{T}_0$ and $\tilde\gamma$ have the same dynamic range---for ensuring numerical stability during NN training.
        We sampled the training set, a validation set for hyperparameter tuning, and a smaller test set
        in the joint $\tilde{\boldsymbol{\pi}} = (\tilde{T}_0, \tilde{\gamma})$ space, all using Sobol' sequences \citep{Sobol_SOBOL196786}. The training set contains 256 distinct parameter labels, the validation set 32, and the test set 16. For each of these labels, we produced 100,000 spectra (from the same 100,000 physical skewer positions across labels) for various training and testing purposes using the exact same procedure as in N24. Figure~\ref{fig:param-space-sample-Planck18} shows a scatter of our training, validation, and test labels. We describe the 
        sampling strategy in more detail in Appendix~\ref{app:sampling}.
        We rescaled all the optical depth $\tau$
        with a constant factor (per distinct thermal model) determined by a linear least-squares fitting in order to match the resulting mean transmission with its observed value of $\bar{F}_\plain{obs} (z=2.2)=0.86$ as measured by \citet{Becker_2013MNRAS.430.2067B}. %

    \subsection{Spectral resolution and noise}\label{sub:noise}
    
        We applied a Gaussian  filter in Fourier space to %
        our spectra corresponding to %
        a spectroscopic resolving power 
        of $R_\plain{FWHM} = \lambda/\Delta\lambda_\mathrm{FWHM} =c/(2\sigma_v\sqrt{2\ln2})=6000$ with the kernel $\mathcal{R}_k = \exp(-k^2 \sigma_v^2/2)$. We additionally applied a Heaviside step function kernel to null the Fourier modes $k>k^*$, $k^*\approx0.094$~s/km corresponding to 
        spectra observable with, e.g., VLT/XSHOOTER.
        We then rebinned the spectra (in real space) with a 16-pixel average to have the final spectrum size of $N_\plain{pixels} = 256$ and each pixel of size $\Delta v_\mathrm{p} = 32.37$ km/s. 

        In this work we consider a simple model of homoscedastic\footnote{The noise level on our pixels $\sigma_\plain{p}$ is fixed within each spectrum.} %
        Gaussian noise in our spectra, $\tilde{F}=F+\epsilon$, $\epsilon \sim \mathbb{N}(0;\sigma_\plain{p})$. %
        (We note that $\sigma_\mathrm{p}$ may vary across spectra in our datasets.) We consider a range of continuum-to-noise ratio (CNR) values between 10 and 500 per 6 km/s, with a uniform prior distribution. This range encompasses the noise levels typical of TPS measurements from targeted quasar samples \citep[e.g.,][]{XQ100_Lya_2017MNRAS.466.4332I, Walther_IGM_2019ApJ...872...13W}. In the Poisson photon counting limit, the noise level $\sigma_\plain{p} = \sigma_6 \sqrt{6/\Delta v_\plain{p}}$ and $\sigma_6 = 1/\plain{CNR}_6$, where $\sigma_6$ and CNR$_6$ correspond to the values for a pixel of size 6 km/s. We note that $\log\sigma_\plain{p} \in [-7.06,-3.14]$.

    \subsection{Traditional summary statistics}
        We consider two widely used traditional summary statistics of the {\lya} forest, the transmission power spectrum (TPS) and the transmission PDF (TPDF). We perform individual and joint analyses of these statistics with noise and spectral resolution effects taken into account and obtain posterior constraints on the parameters $(T_0,\gamma)$ for comparison with our NN machinery.

        The TPS is defined as the variance of the Fourier modes of the transmission contrast field, $\tilde{\delta}_F(k)$, properly normalized. We first measured the TPS from low resolution, noisy spectra (called $\hat{\TPS}_\plain{raw}$) and then applied the following corrections to obtain an estimate $\hat{\TPS}_\plain{pure}$
        of the pure TPS $\TPS_\plain{pure}$, 
        \begin{equation}
            \hat{\TPS}_\plain{pure}(k) = \frac{\hat{\TPS}_\plain{raw} - \hat{\TPS}_\plain{noise}}{ \mathcal{W}_k^2 \mathcal{R}_k^2},
        \end{equation}
        where the noise power spectrum $\hat\TPS_\plain{noise}$
        is measured from $10^6$ random realizations of the noise vector with a known noise level and $\mathcal{W}_k = \sin(k\Delta v_\plain{p}/2) / (k\Delta v_\plain{p}/2)$ is the pixel window function. Since we cut off the modes $k>k^*$, the summary vector $\hat{\TPS}_\plain{pure}$ is of size 122.
        
        \begin{figure*}[h]
            \centering
            \includegraphics[width=0.9\linewidth]{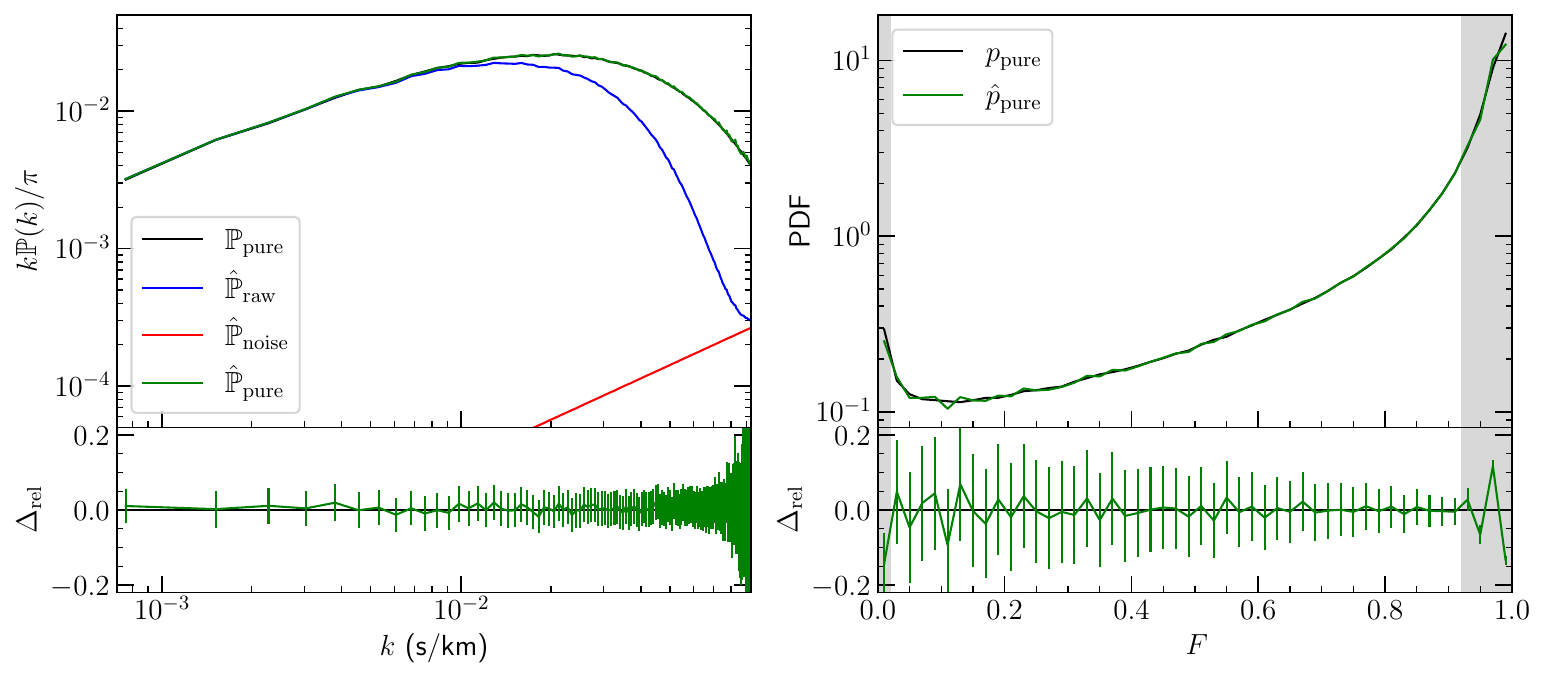}
            \caption{Traditional summary statistics (TPS on the left, TPDF on the right) estimated for the fiducial thermal state of our simulation box and with the mean transmission fixed to its observed value. The raw estimators correspond to a noise level of CNR$_6=30$ ($\sigma_\mathrm{p}=0.014$). The corresponding pure statistics are computed from 100,000 noiseless spectra (with infinite spectral resolution in TPS, with a resolution $R_\mathrm{FWHM}=6000$ for TPDF). The errors correspond to $N_\mathrm{s}=100$ spectra. The gray regions in TPDF correspond to our cuts due to edge effects in the deconvoled estimator.
            }
            \label{fig:traditional-summaries}
        \end{figure*}

        We estimate the TPDF from noisy spectra in the following way. We first compute the PDF $\hat p_\mathrm{raw}$ of the noisy $\tilde{F}$ as a histogram per spectrum in equal width bins ($\Delta F=0.02$) in the range $-0.1\leq \tilde F \leq 1.1$, which is a convolution of the pure PDF $p_\mathrm{pure}$ and $p_\mathrm{noise} \equiv \mathbb{N}(0;\sigma_\plain{p})$. It is normalized so that $\int_{-\infty}^\infty \hat p(F)\,dF = 1$. We then perform Wiener deconvolution \citep{Wiener_1949} of $\hat p_\mathrm{raw}$ and $p_\mathrm{noise}$ to obtain an estimate of the pure PDF, $\hat p_\mathrm{pure}$. Wiener deconvolution aims to estimate a filter $\mathcal{G}$ in the Fourier space such that
        \begin{equation}
            \hat{\mathcal{P}}_\mathrm{pure} = \mathcal{G}\,\cdot\,\hat{\mathcal{P}}_\mathrm{raw} 
            = \frac{\mathcal{N}^*}{|\mathcal{N}|^2+\gamma} \hat{\mathcal{P}}_\mathrm{raw},
        \end{equation}
        where $\mathcal{P,\,G,\,N}$ are quantities in Fourier space and $\gamma$ is a regularization parameter that we tune empirically to $10^{-2}$ to suppress numerical noise amplification. Finally, we crop the PDF to the range $F \in [0.02,0.92]$ for removing edge effects such that the final $\hat{p}_\mathrm{pure}$ vector contains 45 bins. 

        \begin{figure*}
            \centering
            \includegraphics[width=\linewidth]{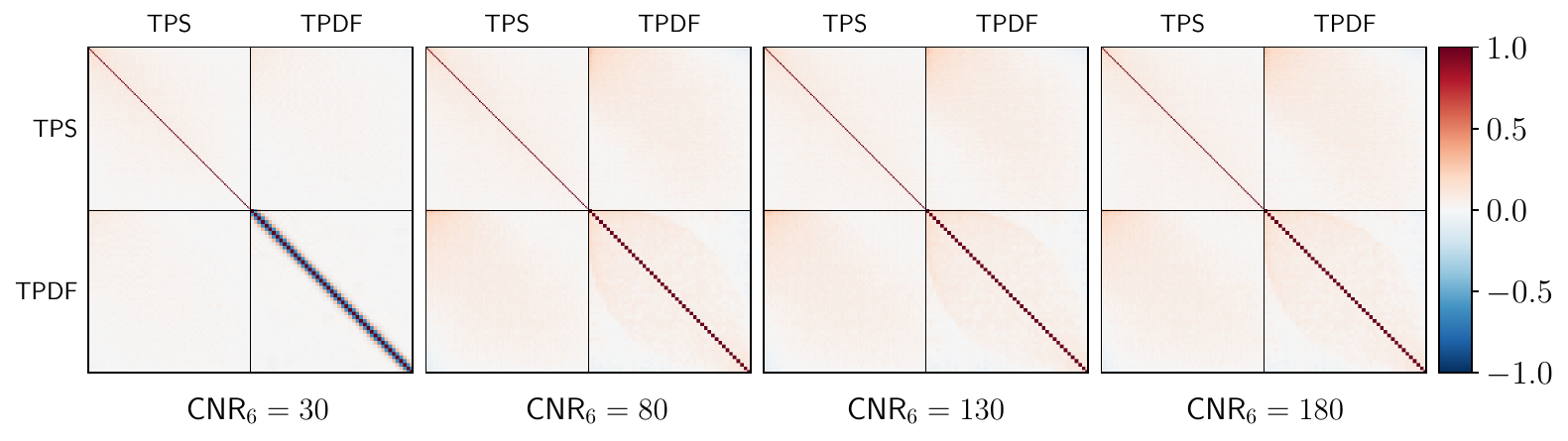}
            \caption{
            The full correlation matrices of the concatenated summary vector for four different noise levels. The TPS is the corrected vector $\hat\TPS_\mathrm{pure}$ with 122 $k$-modes and the TPDF is the deconvolved, cropped $\hat{p}_\mathrm{pure}$ with 45 bins. For each of the four matrices, the two blocks on the principal diagonal are the individual correlation matrices of TPS (top left) and TPDF (bottom right) and the other two blocks show the cross correlation of TPS and TPDF. The different blocks have been resized differently to have an equal area on the plot.
            }
            \label{fig:corr-matrix}
        \end{figure*}
        In Figure~\ref{fig:traditional-summaries} we show our estimates of the TPS (left) and TPDF (right) for our fiducial thermal model and for CNR$_6=30$. For the joint analysis of TPS and TPDF we concatenate the two corrected vectors to have a full summary vector of size 167. We show the full correlation matrices of this joint summary vector (corrected) for four different noise levels in Figure~\ref{fig:corr-matrix}. For a high noise level of CNR$_6=30$, a significant (anti-)correlation among closeby $F$ bins in the TPDF can be seen. This is potentially a result of numerical side effects introduced by the deconvolution with a wide noise PDF. The cross correlations between TPS and TPDF, although mild for all noise levels, are important to account for in the likelihood analysis as found by C25.

\section{NN machinery}\label{sec:machinery}

    \subsection{Architecture}\label{sub:architecture}
        \begin{figure*}[h]
            \centering
            \includegraphics[width=0.9\linewidth]{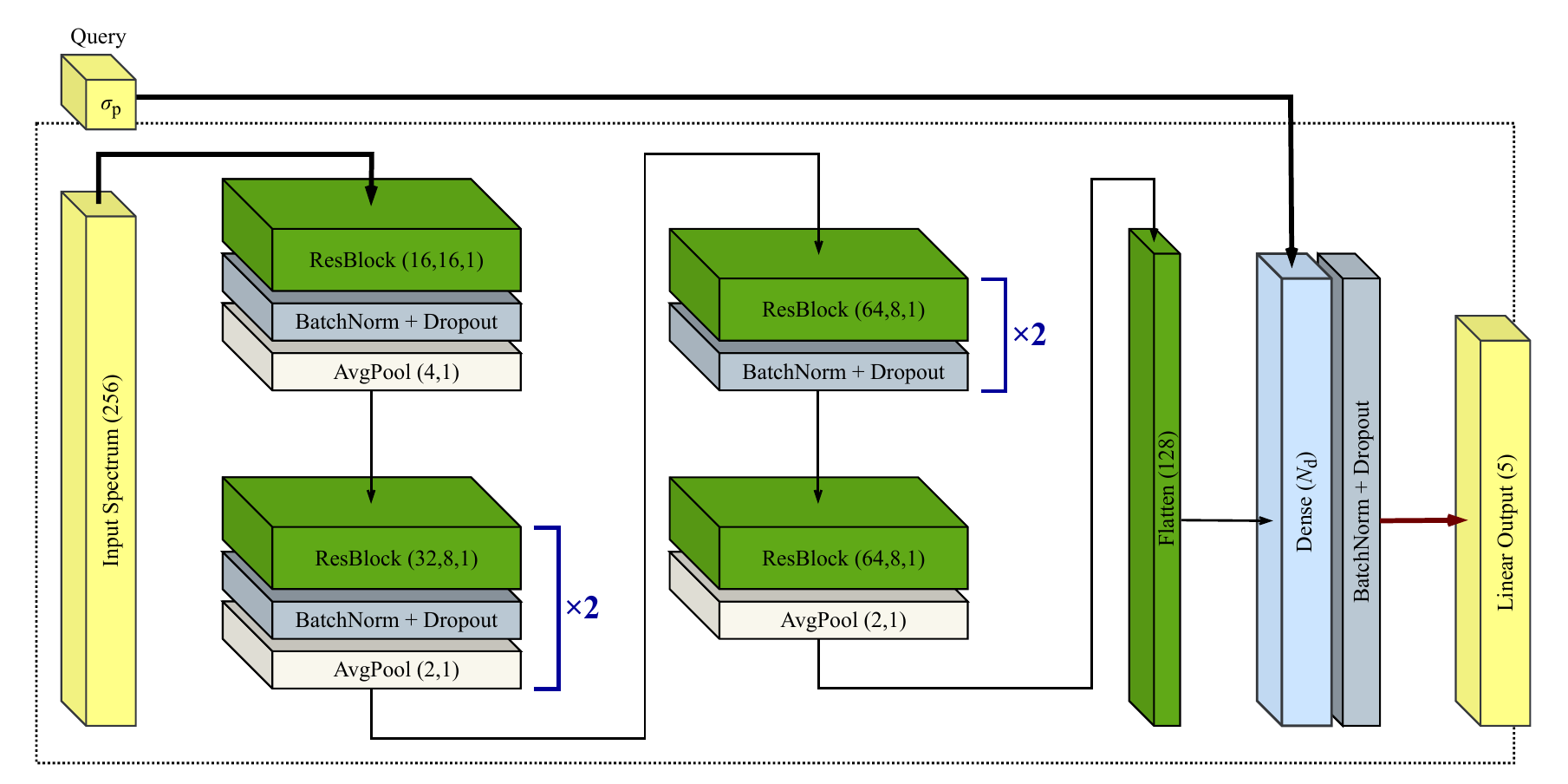}
            \caption{Architecture of {\nsansa}. An input spectrum of size 256 pixels is fed into the network that contains a total of 6 residual blocks and extracts useful features from the field. Batch normalization and dropout are used for regularization and average pooling is used for downsampling. The output of the residual part is flattened, concatenated with the $\sigma_\mathrm{p}$ query, and then fed into a hidden nonlinear layer with $N_\mathrm{d}$ units. Finally, a linear layer with 5 nodes (2 for the summary vector, 3 for its covariance) acts as the output layer. 
            }
            \label{fig:architecture}
        \end{figure*}
        
        Our machinery is based on a 1D ResNet \citep{Resnet_2015arXiv151203385H} convolutional architecture. Over a sequence of six residual blocks similar to the ones in N24, useful features are extracted from noisy input spectra and turned into a flattened feature vector. The (rescaled\footnote{For ensuring numerical stability during training, we use 
        $\hat{\sigma}_\plain{p} \define (2\log\sigma_\plain{p} + 10)/5$ here.
        }) value of $\sigma_\plain{p}$ is concatenated with this vector as a query component and it is then fed into a multilayer perceptron (MLP) of one nonlinear hidden layer with $N_\mathrm{d}=85$ dense nodes and the final linear output layer of 5 nodes for the summary point estimates and the independent components of their covariance matrix. We build this architecture using {\tensorflow}/{\keras} and name it {\nsansa}. It contains 344,952 trainable parameters. We also trained an architecture without the $\sigma_\plain{p}$ query and a different number of hidden units ($N_\mathrm{d}=100$) in the dense nonlinear layer but otherwise identical to {\nsansa}. The detailed architecture of {\nsansa} is shown in Figure~\ref{fig:architecture}. 
        
        We incorporated various regularization techniques into {\nsansa} for promoting generalization (avoiding overfitting to the training set). These are namely, dropout, batch normalization, and L2 kernel- and weight-decay. We used a dropout of $p_1=8.59\times10^{-3}$ after each residual block (except the last) and $p_2=8.63\times10^{-3}$ after the single hidden layer in the MLP (values defined in {\keras} terms). All the convolutional layers in the residual part of the architecture feature a kernel decay with $l_2^\mathrm{\,conv} = 6.61\times10^{-7}$ and the hidden layer in the MLP features a weight decay with $l_2^\mathrm{\,dense} = 5.14\times10^{-6}$. The convolutions have a stride of 2 pixels in the first two residual blocks and a stride of 1 in the rest of them.

    \subsection{Training}\label{sub:training}
        We use the training and validation sets as described in Section~\ref{sec:sim} for training our machine in a supervised way. We use 20,000 spectra per label for training and validation (with fixed skewer positions 
        across labels, different for the two datasets). We additionally augment the training set on the fly with random cyclic permutations and flipping of the spectra as well as randomly drawn noise realizations with properties described in Section~\ref{sub:noise}. We do not augment the validation set, and we add a fixed set of noise realizations to it (that is not varied on the fly), 
        because we would like to compare the state of the network at different epochs of training with the same set of unseen spectra. 

        The training was performed by minimizing the negative log-likelihood loss (NL3) function of the point predictions \emph{w.r.t.} the true labels,
        \begin{equation}
            \mathcal{L}(\tilde{\boldsymbol{\pi}}) = \LogDeterminantSigma + \chi^2,
        \end{equation}
        with $\chi^2 = (\tilde{\boldsymbol{\pi}} - \hat{\boldsymbol{\pi}}) \tilde{\mathbf{\Sigma}}\mathbf{^{-1}} (\tilde{\boldsymbol{\pi}} - \hat{\boldsymbol{\pi}})^T$, where $\hat{[\cdot]}$ are the (rescaled) parameter labels and $\tilde{[\cdot]}$ are quantities output by {\nsansa}. We used the Adam optimizer \citep{Adam_2014arXiv1412.6980K} for this purpose with a fixed learning rate of $\alpha=5\times10^{-4}$. %
        The Adam moments parameters assumed the values $\beta_1 = 0.885$ and $\beta_2=0.999$. 

        \begin{figure}
            \centering
            \includegraphics[width=\linewidth]{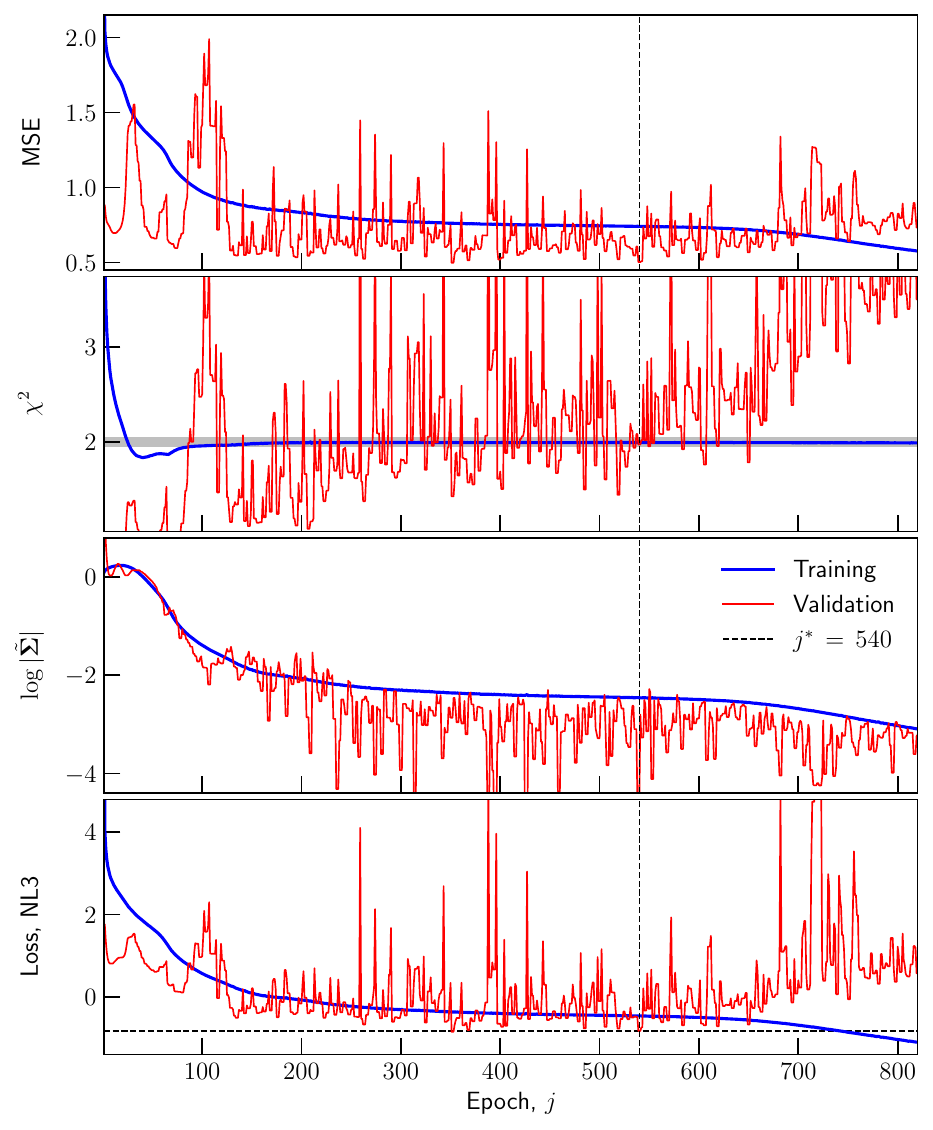}
            \caption{An example of typical learning curves of the {\nsansa} architecture. The gray band in the $\chi^2$ panel indicates our tolerance of $\epsilon=0.05$. Here the minimum of the validation loss is reached at epoch $j^*=540$ and the $\chi^2$ is simultaneously within our tolerance. 
            }
            \label{fig:learning-curves}
        \end{figure}
        We monitor three metrics during training besides the loss value, namely, $\chi^2$, $\LogDeterminantSigma$, and $\mathrm{MSE}=(\tilde{\boldsymbol{\pi}} - \hat{\boldsymbol{\pi}}) (\tilde{\boldsymbol{\pi}} - \hat{\boldsymbol{\pi}})^T$. For a meaningful estimation of $\tilde{\mathbf{\Sigma}}$ by {\nsansa} we expect the average of $\chi^2$ over the training and the validation sets to converge to $N_\plain{params}=2$ as the quality of the network inference improves over the course of training. Figure~\ref{fig:learning-curves}
        shows an example of the learning curves for {\nsansa}. Indeed, $\chi^2$ quickly converges to 2 for the training set and for the validation set it fluctuates close to 2 with slow overall rise. The improvement in the network's state is therefore largely attributed to the minimization of $\LogDeterminantSigma$ or the network's uncertainty in its predictions and of the scatter of the point predictions for any underlying true label. The validation loss $\mathcal{L}_\plain{val}$ improves over the course of the first few hundred epochs and starts getting worse after $\sim600$ epochs, implying overfitting to the training set. We deem the network to have the best possible state for the given hyperparameter set at the epoch $j^*$ at which the validation loss is minimal, $\mathcal{L}_\plain{val} = \mathcal{L}^*_\plain{val},$ and simultaneously $\vert {\chi^*}^2 - 2 \vert \leq \epsilon = 0.05$.

        We performed a Bayesian optimization for tuning the values of the hyperparameters $N_\mathrm{d}$, $l_2^\mathrm{\,conv,\,dense}$, $p_{1,2}$, and Adam $\beta_1$. This procedure is discussed in detail in Appendix~\ref{app:hyperopt}.

    \subsection{Ensemble learning}\label{sub:ensemble}
        NNs are stochastic processes with a large degree of freedom. The choice of hyperparameters, initial weights, random batching of the training data are some of the primary factors that affect the optimal weights of the fully trained machine. In such a scenario, in order to get the best outcome of the infrastructure, we use ensemble learning, which is a powerful tool in machine learning (see, e.g., \citealt{Dietterich_Ensemble_Learning}).
        After finding a set of optimal hyperparameters, we train a committee of 2200 NNs with varying initial weights and on-the-fly stochasticity. We then employ bootstrap aggregation (``bagging,'' \citealt{Breiman_Bagging}) to combine the individual network predictions in form of an averaging weighted by $1/\sqrt{|\tilde{\mathbf{\Sigma}}|}$. The bagged results of the committee are expected to be statistically more stable than individual networks \textit{w.r.t.} variations in the input vectors. 
         In the following, the predictions of the machine are assumed to be the bagged results of the committee unless otherwise specified.

\section{Inference}\label{sec:inference-method}
    Our model {\nsansa} acts as a compression of the data into optimal summary statistics of the same size as the number of parameters, $N_\plain{params}=2$. The resulting summary vector $\mathbf{\mathcal{\hat{S}}}$ can be used to infer the parameters $(T_0,\gamma)$. We explore two different methods for estimating the posterior constraints with {\nsansa} in this work: (i) Gaussian likelihood analysis and (ii) density estimation likelihood free inference (DELFI). 

    \subsection{Gaussian likelihood}\label{sub:gaussian-likelihood}
        In the likelihood case, we chose a multivariate Gaussian likelihood function of the NN-compressed summary vector $\mathcal{S}$ (as well as the traditional summaries TPS, TPDF and their combination) of the form 
        \begin{equation}
            \log L_{N_\mathrm{s}}(\mathbf{\boldsymbol{\pi}}) 
            \sim -\frac{1}{2}\log \Det{\mathbf{C}_{N_\mathrm{s}}} + \frac{(\bar{\mathcal{S}}_{N_\mathrm{s}} - \boldsymbol{\boldsymbol{\mu}}) \mathbf{C}_{N_\mathrm{s}}\mathbf{^{-1}} (\bar{\mathcal{S}}_{N_\mathrm{s}} - \boldsymbol{\mu})^T}{2},
        \end{equation}
        where $N_\mathrm{s}$ refers to the number of sight lines the amount of information corresponds to, $\mathbf{C}$ %
        is the covariance matrix of $\mathcal{S}$, $\bar{\mathcal{S}}$ refers to the average summary vector for any given input data, and $\boldsymbol{\mu} \equiv \boldsymbol{\mu}(\boldsymbol{\pi}; \sigma_\plain{p})$ is our model (emulator) for this summary statistic. 
        
        We built an emulator of the {\nsansa} statistic with a linear interpolation in the 3D space of ($\boldsymbol{\pi},\sigma_\mathrm{p}$) using the mean predictions of {\nsansa} from 20,000 previously unseen spectra per distinct $\hat{\boldsymbol{\pi}}$ of the joint training and validation sets. (These correspond to the same skewer positions as those used for validation.) For the traditional summaries we created emulators via cubic spline interpolation of $k$-modes (for TPS) and $F$ bins (for TPDF) independently using the respective pure estimators over the training labels, computed using 100,000 spectra per label.
    
        We compute the necessary covariance matrices via bootstrapping in the following way. We estimate mean summary vectors $\{\mathbf{s}_i\}$ over mutually exclusive subsets of size $N_\mathrm{s}=100$, exhausting the underlying test set (described in more detail in Section~\ref{sec:results}). 
        We then repeat this stratified subsampling 1,000 times by introducing randomness in selecting the subsets.
        The covariance is then estimated by 
        \begin{equation}
            \mathbf{C_s} = \frac{1}{N-1} \sum_{i=1}^{N}(\mathbf{s}_i - \mathbf{\bar{s}})^T (\mathbf{s}_i - \mathbf{\bar{s}}),
        \end{equation}
        where $N$ is the total number of subsets and $\mathbf{\bar{s}}$ is the global mean summary vector. The errors thus correspond to the information content of $N_\mathrm{s}=100$ spectra (or equivalently a {\lya} forest of length 8 cGpc$/h$).

    \subsection{Density estimation likelihood free inference (DELFI)}\label{subsub:lfi}
         LFI, as the name indicates, gets rid of any assumptions about the form of the likelihood function for Bayesian inference. In this work we employed a popular approach to LFI, namely, density estimation likelihood free inference \citep[DELFI; e.g.,][]{DELFI_2018MNRAS.477.2874A} with {\nsansa}. (It is important to note that we do not perform DELFI with the traditional summaries here.) The posterior is expressed in terms of the joint distribution of summary vectors and parameters evaluated at the summary values of the underlying data, 
        \begin{equation}
            p(\boldsymbol{\pi} \vert \mathbf{d=d_*}) \propto p(\mathbf{d_*}\vert\boldsymbol{\pi})p(\boldsymbol{\pi}) =  p_\plain{joint}(\boldsymbol{\pi},\mathbf{d}) \big\vert_{\mathbf{d=d_*}}.
        \end{equation}

        We use a Gaussian mixture model (GMM) as a surrogate for the estimation of the joint density $p_\mathrm{joint}(\boldsymbol{\pi},\mathbf{d})$. We treat 
        $\sigma_\plain{p}$ 
        of each spectrum as part of the data vector such that $\mathbf{d} \define (\mathcal{\hat{S}, \sigma_\plain{p}})$. Hence, the size of the joint vector for density estimation, $(\boldsymbol{\pi},\mathbf{d})$, is 5. We created a large dataset for training the GMM surrogate as follows. We first selected all the training and validation labels within our prior range (as indicated by the gray region in Figure~\ref{fig:param-space-sample-Planck18}) amounting to 177 and obtained {\nsansa} predictions for 10,000 spectra per label in that joint set (a subset of those used for covariance estimation in the Gaussian likelihood case). We iterated this procedure for 19 different values of CNR$_6$ linearly spaced between 20 and 200. For each distinct CNR$_6$, we generated a large set of mean summary vectors over subsets of $N_\mathrm{s} = 100$ spectra with 1,000 repetitions of the stratified subsampling as described in Section~\ref{sub:gaussian-likelihood}, thus completing the GMM training set generation. The size of this dataset is 336,300,000 joint $(\boldsymbol{\pi},\mathbf{d})$ vectors.
        
        We used the {\python} package \textsc{scikit-learn} for fitting the GMM surrogate, with a regularization factor of $10^{-5}$ added to the diagonal of the component covariance matrices to ensure numerical sanity, stability, and prevent overfitting, the latter of which was asserted through a convergence analysis. We fit 30 GMMs through our collective dataset, each with a unique number of mixture components $K=$ 1 through 30. 
        Based on the results of this convergence analysis, we empirically chose the GMM surrogate with $K=23$ for inference with DELFI.

    \begin{figure*}[h]
        \centering
        \includegraphics[width=0.9\linewidth]{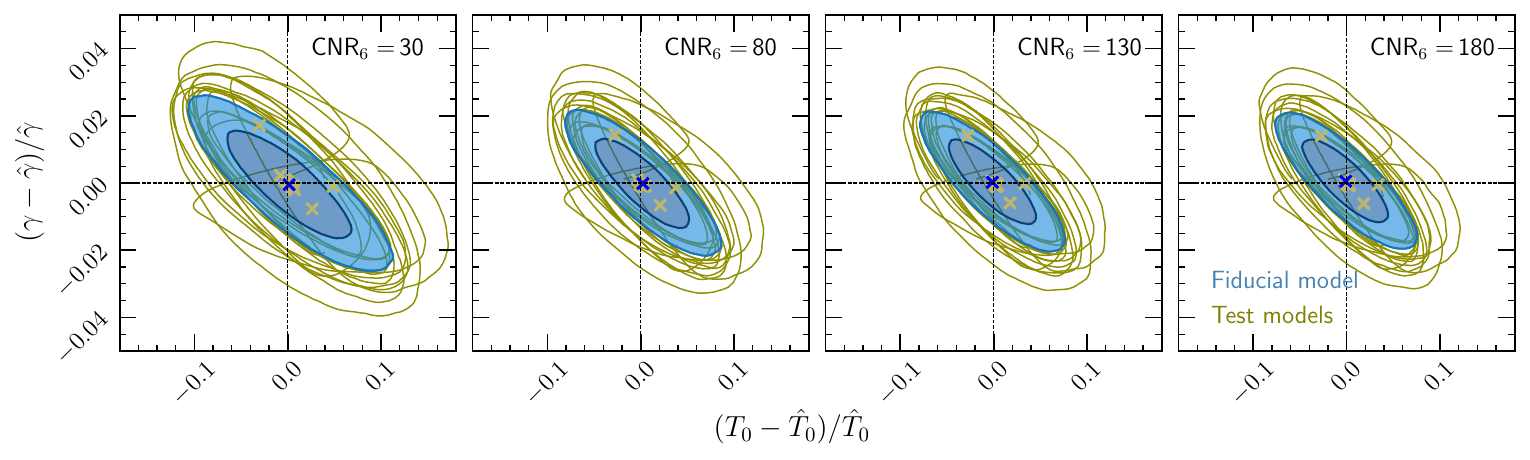}
        \caption{A comparison of posterior constraints on $(\boldsymbol{\pi}-\hat{\boldsymbol{\pi}})/\hat{\boldsymbol{\pi}}$ using {\nsansa} likelihood among 16 different test $\hat{\boldsymbol{\pi}}$ models and the fiducial model for four different noise levels. For the test models only the 95\% credibility contours are shown, whereas for the fiducial model the 68\% and 95\% credibility contours are shown. For all of them the posterior means are also shown with crosses of the corresponding colors.}
        \label{fig:nsansa-post-all-test-lkd}
    \end{figure*}
    \begin{figure}
        \centering
        \includegraphics[width=\linewidth]{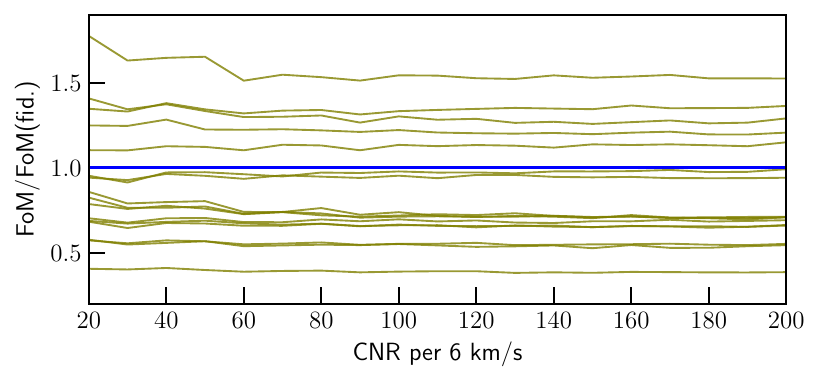}
        \caption{The ratio of the FoM between the test models and the fiducial model against the noise level for the {\nsansa} likelihood.}
        \label{fig:fom-ratio-test-to-fid}
    \end{figure}    

\section{Results and discussion}\label{sec:results}
    \begin{figure*}
        \centering
        \includegraphics[width=0.9\linewidth]{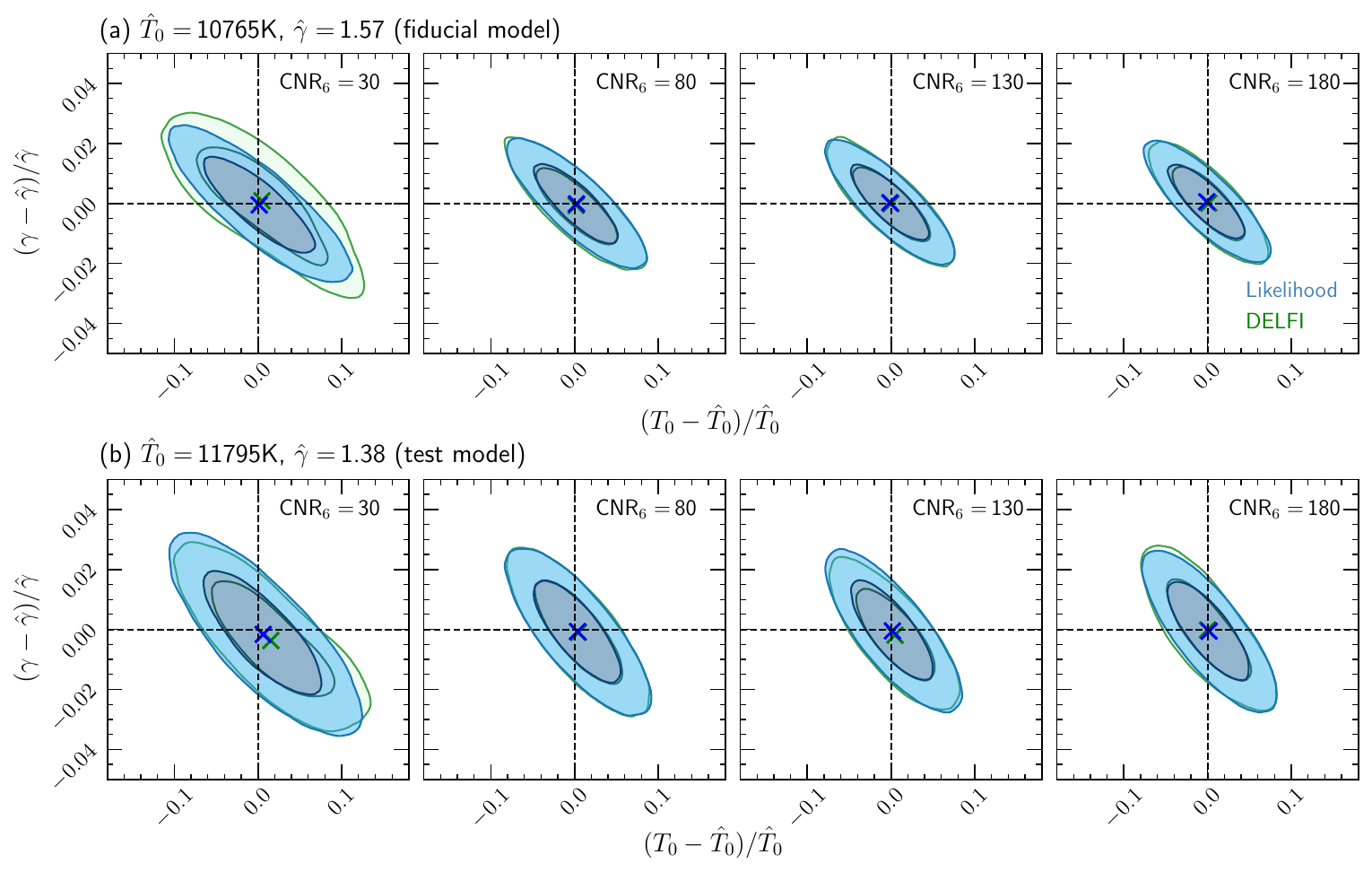}
        \caption{A comparison of {\nsansa} Gaussian likelihood analysis and DELFI in terms of posterior constraints on $(\boldsymbol{\pi}-\hat{\boldsymbol{\pi}})/\hat{\boldsymbol{\pi}}$ for the fiducial TDR model (top) and a test model (bottom) for four different noise levels. In each case, a cross marker of the corresponding color indicates the posterior mean. For DELFI, a GMM of $K=23$ components was used to estimate the joint $(\boldsymbol{\pi},\mathbf{d})$ density.}
        \label{fig:nsansa-post-delfi}
    \end{figure*} 
    
    \begin{figure}[h]
        \centering
        \includegraphics[width=\linewidth]{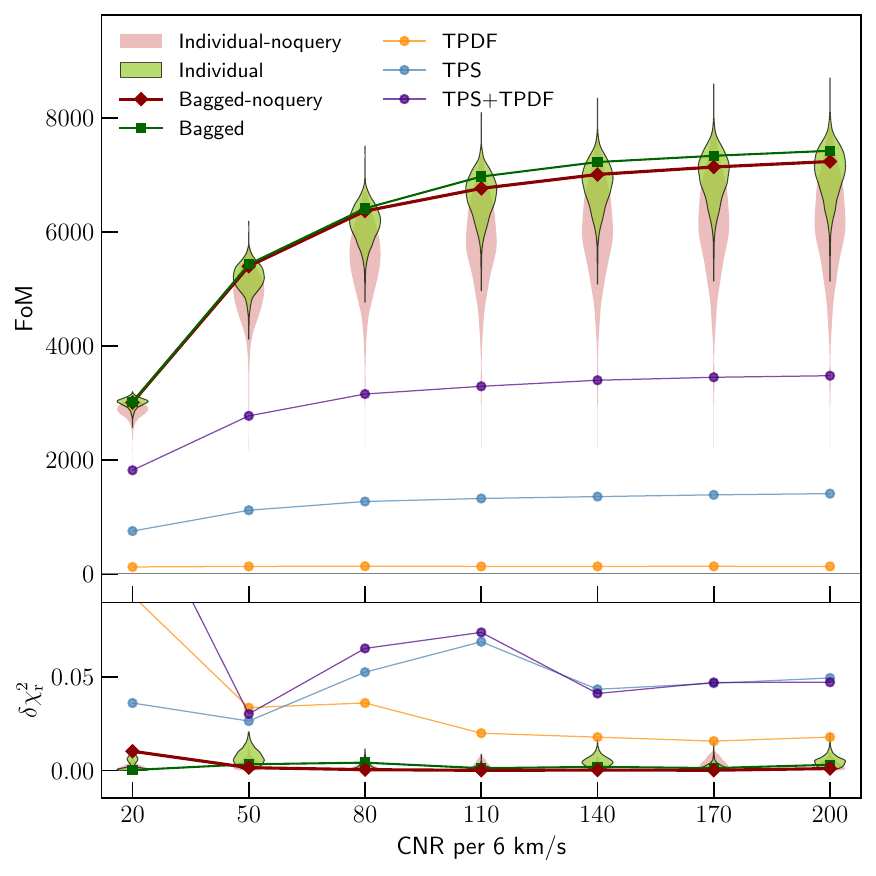}
        \caption{Posterior FoM (top) and $\delchisq$ (bottom) against CNR for {\nsansa} and an equivalent architecture without the $\sigma_\mathrm{p}$ query, compared to the TPS, TPDF and joint TPS+TPDF constraints for the fiducial thermal model. All values correspond to the Gaussian likelihood inference. The violins show the distributions of the metrics of the individual members of the ensemble(s).
        }
        \label{fig:fom_cnr_nsansa_pspdf}
    \end{figure}
    We performed tests of our machinery for a set of different noise levels and determined the behavior of the inference outcome \textit{w.r.t.} CNR. Our full test dataset contains 10,000 spectra from each of the 16 test models (labels) described in Section~\ref{sub:hydro-sim} as well as the fiducial thermal model. These spectra are generated using a completely different set of skewers than those used for training, validation, and model-building.
    We additionally sampled 19 linearly spaced CNR$_6$ values between 20 and 200 in total for various test purposes\footnote{This is a subset of the full range of CNR$_6$ used during training to ensure that tests fall well inside our training prior boundaries.}. For each distinct value of CNR$_6$ in our sample, we added homoscedastic Gaussian noise 
    with the corresponding $\sigma_\mathrm{p}$ to the test spectra, all with random independent phases across pixels, spectra, and labels. We then fed those into {\nsansa} and obtained a set of (bagged) predictions that we treat as summary vectors and performed Bayesian inference with them, with a Gaussian likelihood as well as DELFI. For a comparison, we also performed inference with the TPS and TPDF (combined) estimated with the same set of test spectra for all the sampled noise levels. In each case, we sampled from the posterior distribution of $(\boldsymbol{\pi} - \hat{\boldsymbol{\pi}})/\hat{\boldsymbol{\pi}}$ using MCMC with affine invariant sampling\footnote{
    Performed using the {\python} package \texttt{emcee}, \href{https://emcee.readthedocs.io/en/stable/}{https://emcee.readthedocs.io}.}. 

    \begin{figure*}
        \centering
        \includegraphics[width=0.9\linewidth]{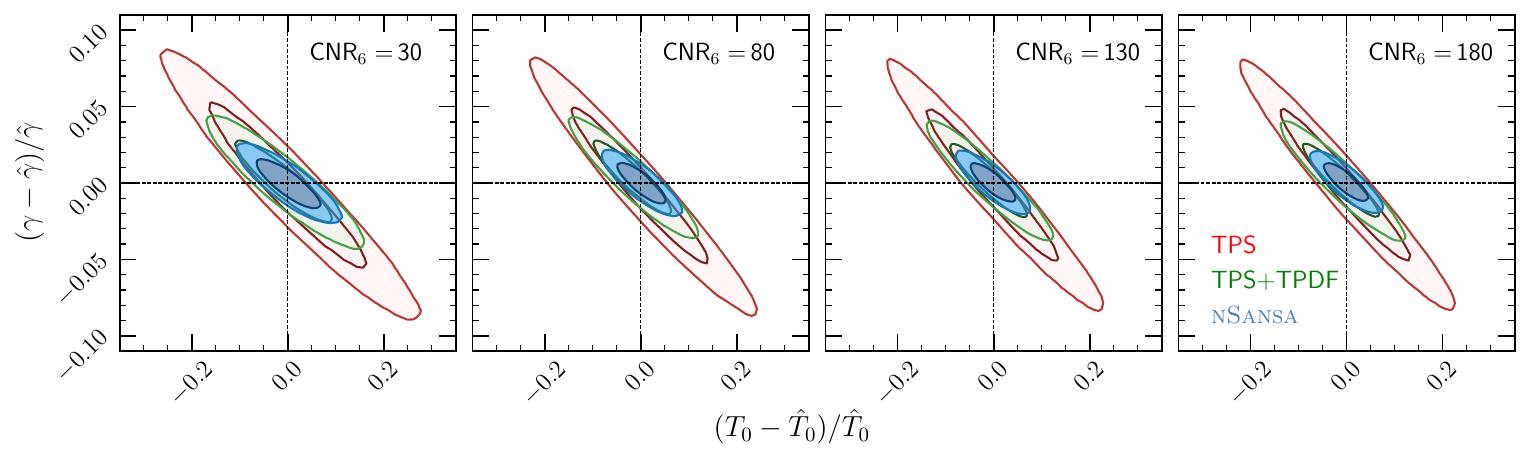}
        \caption{Posterior constraints on $(\boldsymbol{\pi}-\hat{\boldsymbol{\pi}})/\hat{\boldsymbol{\pi}}$ using {\nsansa} for four different noise levels compared to the constraints by TPS and TPS+TPDF. A Gaussian likelihood was used for inference in all these cases.} 
        \label{fig:nsansa-post-vs-pspdf}
    \end{figure*}
    
    We characterize the precision and accuracy of our posterior constraints using the following two metrics: 
    \begin{enumerate}[(i)]
        \item We employ the reciprocal of the contour size as our figure of merit (FoM) of the posterior precision, such that 
            \begin{equation}
                \plain{FoM} \define 1/\Det{\mathbf{C}_\plain{post}}^{1/N_\plain{params}},
            \end{equation}
            where the covariance $\mathbf{C}_\mathrm{post}$ is estimated from the posterior MCMC sample. A larger FoM implies tighter constraints.

        \item We quantify the accuracy of our constraints with a reduced $\chi^2$ metric, 
            \begin{equation}
                \delchisq = \langle \chi^2\rangle/N_\plain{params} - 1,
            \end{equation}
            where $\chi^2 = (\boldsymbol{\pi} - \hat{\boldsymbol{\pi}})\mathbf{C}_\mathrm{post}^\mathbf{-1} (\boldsymbol{\pi} - \hat{\boldsymbol{\pi}})^T$ 
            and the average is taken over the MCMC posterior sample. A $\delchisq$ of 0 implies that the mean of the posterior exactly recovers the truth 
            and $\delchisq$ of 2 implies that the truth is $2\sigma$ away from the mean of the posterior.
    \end{enumerate}
    
    We first show a comparison of the posterior constraints using the {\nsansa} likelihood for all the 16 test labels $\hat{\boldsymbol{\pi}}$ as well as the fiducial model for four different noise levels in Figure~\ref{fig:nsansa-post-all-test-lkd}. The contours for all the test models (except three models at the edges of our prior) show a statistically similar behavior, lending credibility to the claim that the performance of {\nsansa} is stable across our prior in $(\boldsymbol{\pi},\sigma_\mathrm{p})$. This is further evidenced by the ratios of the FoM between those test models and the fiducial model as shown in Figure~\ref{fig:fom-ratio-test-to-fid}.

    We then compare the {\nsansa} posterior constraints between the likelihood and the DELFI cases in Figure~\ref{fig:nsansa-post-delfi}. The top panel thereof shows the posterior contours for our fiducial thermal model and the bottom panel a test model of the 16. The contours in the two cases agree with each other to an excellent degree for both the models at most noise levels shown. The DELFI precision evolves with the CNR very similarly to the likelihood case and the DELFI contours only exhibit a marginally non-Gaussian behavior at CNR$_6=30$. This demonstrates that a Gaussian function is a reasonable approximation of the likelihood for {\nsansa} over the entire prior range considered in this work. 

    In Figure~\ref{fig:fom_cnr_nsansa_pspdf} we show our posterior metrics against noise level for {\nsansa} as well as the traditional summary statistics considered in this work, and in Figure~\ref{fig:nsansa-post-vs-pspdf} we show the posterior constraints for the fiducial model for four different noise levels using {\nsansa} as well as the TPS and TPS+TPDF. This comparison is for our fiducial thermal model and the Gaussian likelihood case for all the summaries. At all noise levels the constraints by {\nsansa} are more precise and more accurate than the traditional summaries and the boost in precision increases gradually with the CNR. In particular, the {\nsansa} FoM is 1.65 times larger than the TPS+TPDF case at CNR$_6=20$ and 2.12 times larger at CNR$_6=200$. At CNR$_6>100$, the original architecture with a $\sigma_\mathrm{p}$-query performs slightly better than its non-query counterpart in terms of precision for the bagged results of the ensemble. At all CNR the FoM distribution of the ensemble with the query generally centers around higher FoM than the non-query ensemble. All members in the query ensemble have a higher FoM than the TPS+TPDF combination, whereas for the non-query ensemble a tail extends to lower FoM, rendering it an inefficient architecture to be used standalone (without ensembles). It is also noteworthy that in our findings the best-performing member of a committee at one noise level is not necessarily so at another, further raising the issue of stability of the inference using a single member network. Nonetheless, the bagged results of the non-query version are especially interesting since, unlike its query counterpart or the traditional summaries, no information of the underlying noise level needs to be supplied whatsoever.

\section{Conclusion and outlook}\label{sec:conclusion}
    We extended the proof-of-concept analysis of N24 by incorporating certain real-world nuisance effects in the {\lya} forest observations to make a deep learning field-level inference pipeline such as {\lyanna} more reliable for actual data. 
    Namely, we studied the impact of instrumental noise in the spectra on the precision and accuracy of the NN inference. We constructed a 1D ResNet convolutional architecture called {\nsansa} to recover the power law TDR parameters $(T_0,\gamma)$ of the IGM from  medium resolution,
    noisy {\lya} forest data akin to \citet{XQ100_2016A&A...594A..91L}. We trained this machinery with a large set of labeled mock spectra generated from hydrodynamical simulations, including a range of noise levels characteristic of targeted, individual spectroscopic observations. We performed an extensive hyperparameter tuning based on Bayesian optimization using a completely disjoint validation set and thereafter trained a committee of 2200 networks with the best hyperparameter values for ensuring statistical stability of the results via ensemble learning. The resultant pipeline can be treated as an optimal compression of the {\lya} transmission field into a summary statistic that is degenerate with the target parameters of interest. 
    
    We performed Bayesian inference with {\nsansa} using a Gaussian likelihood (as is common practice) as well as without any assumptions of the form of the likelihood.
    In the latter case we estimated the joint density of the parameters (labels) and the NN-compressed summaries via a GMM surrogate for DELFI. Finally, we conducted a detailed investigation of this inference framework with a previously unseen realistic mock test dataset and characterized its behavior with varying noise levels on the spectra. When compared to traditional summary statistics of the field such as TPS and TPDF---each of them having been corrected for noise---our machinery exhibits enhanced precision and accuracy of inference, which suggests that it can extract useful non-Gaussian features of the transmission
    field that are not captured by the traditional summaries even when they are buried under noise.

    We compared the Gaussian likelihood inference and DELFI against each other for {\nsansa} to draw interesting conclusions about the NN-compressed summary statistic. The posterior precision for the two is comparable in the case of a GMM with 23 mixture components for DELFI and only mild deviations from a Gaussian posterior appear. In light of the computational resources required to build a 5D surrogate density and the eventual time of running MCMC with it, a Gaussian (distribution) 
    can be regarded as a reasonable approximation of the form of likelihood for {\nsansa}. We discourage a comparison of {\nsansa} DELFI with the constraints of traditional summaries (TPS+TPDF) performed in this work at this point, since an equivalent likelihood-free inference framework for the latter is lacking as of yet, and we motivate future research in this direction to address the compatibility of simpler likelihood assumptions for TPS and TPDF.

    It is instructive to note that the improvement in posterior precision over TPS+TPDF we found in this work (over the whole CNR$_6$ range studied) is not as large as in the pure, noise-free case of N24. This could partially stem from the fact that the traditional summary vectors are corrected for noise using known analytical properties (a subtraction of noise power for the TPS and a deconvolution of the noise PDF for the TPDF). On the other hand, supplying the information of $\sigma_\mathrm{p}$ in the {\nsansa} workflow via a query only leads to a marginal gain in precision---only at high CNR---over an equivalent pipeline without the said query, signaling to the inability of the {\nsansa} architecture to meaningfully utilize this auxiliary information of the noise level.
    Further investigation may potentially reveal alternative ways to account for this information into a field-level inference pipeline to yield a correction of the similar degree as for the human defined summaries. 

     Moreover, in this work we focused solely on two realistic nuisance effects, namely noise and spectral resolution. While relatively straightforward to handle with traditional summaries, they are two of the biggest concerns for spectrum-based inference from targeted spectroscopic observations of quasars
    and thus merit a methodological investigation such as this into the usefulness of inference with deep learning. However, they are accompanied by a set of other modeling and physical systematic effects, especially in large cosmological survey datasets. These are, for instance, quasar continuum fitting uncertainties, metal absorption lines, damped {\lya} absorbers (DLAs), etc. These must be adequately incorporated into the next generation field-level frameworks for robust and reliable inference with large real-world datasets using artificial intelligence. 

\section*{Data Availability}
    The data and/or code pertaining to the analyses carried out in this paper shall be made available upon reasonable request to the corresponding author.

\begin{acknowledgements}
We thank the members of the chair of Astrophysics, Cosmology, and Artificial Intelligence (ACAI) at LMU Munich for useful discussions and support. We acknowledge the Faculty of Physics of LMU Munich as well as the Leibniz Rechenzentrum (LRZ) via the MCML initiative for providing invaluable computing resources and user support for this project.
We acknowledge PRACE for awarding us access to Joliot-Curie at GENCI@CEA, France via proposal 2019204900. We also acknowledge support from the Excellence Cluster ORIGINS which is funded by the Deutsche Forschungsgemeinschaft (DFG, German Research Foundation) under Germany’s Excellence Strategy – EXC-2094 – 390783311. PN thanks the German Academic Exchange Service (DAAD) for providing a scholarship to carry out this research. MW acknowledges support by the project AIM@LMU funded by the German Federal Ministry of Education and Research (BMBF) under the grant number 16DHBKI013.
\end{acknowledgements}

  \bibliographystyle{aa} %
  \bibliography{main} %

\appendix

\section{Parameter space sampling}\label{app:sampling}
    We sampled our training, validation, and test labels in the parameter space using Sobol' sequences for an optimal prior volume filling. In the orthogonal space $(\alpha,\beta)$ of the TDR parameters $(T_0,\gamma)$ as discussed in Appendix B of N24, we first created a sample of 256 labels of our training set in the range $-1.1<\alpha<1.1$ and $-1.5<\beta<1.5$. We then applied rescalings of the labels outside our prior, $-1\leq\alpha,\beta\leq 1$, as shown in Figure~\ref{fig:param-space-sample-Planck18} (gray region), such that the 2d density 
    of labels in that region is $1/4$th of the original. This is motivated by our observations in Appendix C of N24. (It is noteworthy that, unlike N24, we did not apply a linear transformation to the predictions by {\nsansa} after training and used the raw predictions throughout all our analyses. Any biases due to the limited range in sampling are automatically taken into account while creating an emulator for the likelihood analysis and a GMM surrogate for DELFI.) We sampled the validation and test sets both entirely within our prior range.

\section{Hyperparameter optimization}\label{app:hyperopt}
    After manually optimizing the overall skeleton of the {\nsansa} architecture, we proceeded to tune some of the more refined hyperparameters of the system. Our objective entailed finding values of hyperparameters $\theta$ that minimize $\score(\theta)$, i.e.,
    $\theta_\plain{opt} = \arg\!\min_\theta \score(\theta)$. We used the open-source {\python} package {\optuna} for tuning the hyperparameters $\beta_1$ (Adam), dropout rates and the $l_2$ amplitudes of regularization of convolutional kernels and MLP layers, and the number of nodes in the hidden layer of the MLP. Our search space priors are listed in Table~\ref{tab:search-space}.
    \begin{table}[h]
        \caption{The search space for our hyperparameter optimization}
        \label{tab:search-space}
        \centering
        \begin{tabular}{lll}
            \toprule
            \textbf{Hyperparameter}                 & \textbf{Range}        & \textbf{Sampling} \\
            \midrule
            Adam $\beta_1$                          & $[0.88,0.92]$         & (unif. in linear space) \\
            Kernel decay, $l_2^\mathrm{\,conv}$     & $[10^{-8},10^{-4}]$   & (unif. in log space) \\
            Weight decay, $l_2^\mathrm{\,dense}$    & $[10^{-8},10^{-4}]$   & (unif. in log space) \\
            Dropout (conv.), $p_1$                  & $[0.0,0.2]$           & (unif. in linear space) \\
            Dropout (MLP), $p_2$                    & $[0.0,0.1]$           & (unif. in linear space) \\
            \# hidden nodes, $N_\mathrm{d}$         & $[75,100]$            & (unif. over integers) \\
            \bottomrule
        \end{tabular}
    \end{table}
    We employed Bayesian optimization based on Gaussian processes (GP) with the \texttt{GPSampler} class of {\optuna}. We initially ran 128 trials with hyperparameters sampled with a quasi Monte Carlo (QMC) strategy based on Sobol' sequences (\texttt{QMCSampler}) and then built and refined our GP surrogate model to sample hyperparameter sets for further trials. We distinguish the two phases as the QMC (initial phase) and PSC (post surrogate construction phase). {\optuna} uses log of expected improvement as the acquisition function by default. We refer interested readers to the {\optuna} documentation for more details (\href{https://optuna.readthedocs.io/en/stable/reference/samplers/generated/optuna.samplers.GPSampler.html}{https://optuna.readthedocs.io}).
    \begin{figure}
        \centering
        \includegraphics[width=\columnwidth]{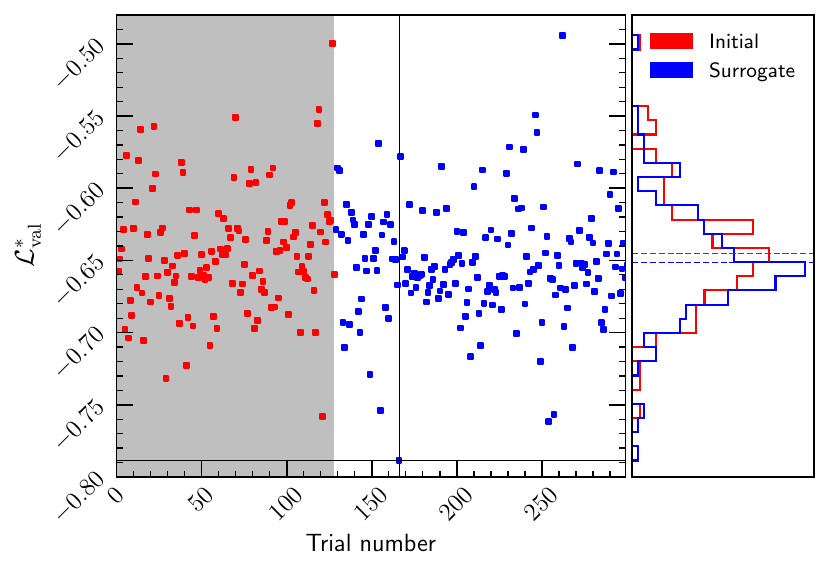}
        \caption{\textbf{Left}: Minimal validation loss against trial number for our Bayesian hyperparameter tuning. The gray shaded region indicates the QMC phase to build the surrogate out of. \textbf{Right}: Normalized histograms of the minimal validation loss across trials in our hyperparameter optimization. We show the QMC and the PSC phases separately. The dashed vertical lines show the mean loss values in both the phases. The PSC phase leans slightly toward smaller loss values than the QMC phase. 
        }
        \label{fig:trials-loss-scatter-hists}
    \end{figure}
    
    Figure~\ref{fig:trials-loss-scatter-hists} 
    shows a scatter of the minimum validation loss obtained, $\mathcal{L}^*_\plain{val}$, as a function of trial number as well as histograms of $\mathcal{L}^*_\plain{val}$. The best value of $\mathcal{L}^*_\plain{val}$ is found in the PSC phase at trial 165+1. The distribution of $\mathcal{L}^*_\plain{val}$ leans only \emph{slightly} toward smaller values in the PSC phase than in QMC contrary to our expectation of an informed Bayesian optimization algorithm, hinting at the inefficacy of the acquisition function and the surrogate generation internal to {\optuna}. Nevertheless, the best $\mathcal{L}^*_\plain{val}$ is a $>2\sigma$ outlier in both the distributions.

\end{document}